# Multiple spatial and wavelength conversion operations based on a frequency-degenerated intermodal four-wave-mixing process in a graded-index 6-LP few mode fiber

## H. ZHANG,[1,2] M. BIGOT-ASTRUC,[3] P. SILLARD,[3] G. MILLOT,[1] B. KIBLER[1], AND J. FATOME[1,4,5]*

[1] Laboratoire Interdisciplinaire Carnot de Bourgogne, UMR 6303 CNRS-Université Bourgogne Franche-Comté, Dijon, France
[2] Extreme Optoelectromechanix Laboratory, School of Physics and Electronic Sciences, East China Normal University, Shanghai, 200241, China
[3] Prysmian Group, Parc des Industries Artois Flandres, Haisnes 62092, France
[4] Department of Physics, The University of Auckland, Private Bag 92019, Auckland 1142, New Zealand
[5] The Dodd-Walls Centre for Photonic and Quantum Technologies, New Zealand
*Corresponding author: Julien.Fatome@u-bourgogne.fr

We report on the experimental observation of a simultaneous threefold wavelength and spatial conversion process at telecommunication wavelengths taking place in a 6-LP-mode graded-index few-mode fiber. The physical mechanism is based on parallel and phase-matched frequency-degenerated inter-modal four-wave mixing (FD-IFWM) phenomena occurring between the fundamental mode and higher-order spatial modes. More precisely, a single high-power frequency-degenerated pump wave is simultaneously injected in the four spatial modes $LP_{01}$, $LP_{11}$, $LP_{02}$ and $LP_{31}$ of a 1.8-km long graded-index few-mode fiber together with three independent signals in the fundamental mode. By means of three parallel phase-matched FD-IFWM interactions, these initial signals are then simultaneously spatially and frequency converted from the fundamental mode to specific high-order modes. The influence of the differential modal group delay is also investigated and shows that the walk-off between the spatially multiplexed signals significantly limits the bandwidth of the conversion process for telecom applications.

## 1. INTRODUCTION

In the last few years, space-division multiplexing (SDM) has emerged as a promising technology in optical communications to match the continuously increasing data traffic in fiber networks and thus prevent potential "data-bottleneck" in the next decades [1-7]. To take over from the classical single-mode platform, multicore fibers and few-mode fibers (FMFs) have emerged as a potential solution to fulfil the growing demand of data traffic [1-2]. Thanks to this further degree-of-freedom, each fiber core or mode can then operate as an additional multiplexed spatial channel, thus enabling to largely overpass the capacity limits of current single-mode systems [3-5]. In this context, the developments of novel spatial components as well as specifically designed FMFs have stimulated the discovery of new nonlinear effects taking place in multimode fibers [8-17] and especially for nonlinear signal processing based on inter-modal four-wave mixing (IFWM) [18-29]. In fact, the difference of propagation constant between spatially multiplexed signals offers an additional degree-of-freedom for phase-matching condition compared to the classical single-mode configuration. Consequently, large frequency-shift four-wave mixing interactions can be obtained without the restriction to operate near the zero-dispersion wavelength as in standard single-mode fibers [30-32]. In this new contribution, we explore a particular form of spatially non-degenerated but frequency-degenerated pumped IFWM, called FD-IFWM, for which a single pump wavelength around 1.55 μm is involved into three simultaneous spatial and wavelength conversion processes. This FD-IFWM mechanism is based on the spatial multiplexing of a single high-power continuous pump wave (frequency-degeneration of the process) injected simultaneously into four distinct spatial modes (spatially non-degenerated) of a 1.8-km long graded-index few-mode fiber, respectively $LP_{01}$, $LP_{11}$, $LP_{02}$ and $LP_{13}$. Three additional independent signals are then combined into the fundamental mode whose wavelengths are specifically selected to each fulfill a phase-matching condition of FD-IFWM between $LP_{01}$ and one of the other high-order modes understudy. As a result, the three initial signals are independently frequency and spatially converted from the fundamental mode towards a specifically targeted high-order mode. Compare to previous works, especially our ref. [23], the present results highlight for the first time of our knowledge, a single-pump configuration enabling a three-fold simultaneous wavelength and spatial conversion process of three telecom signals in FMF. This novel configuration appears much simpler that previous works reported in refs. [22-23, 27] that exploited two different pump waves to satisfy the phase matching condition. Furthermore, the influence of the differential modal group delay (DMGD) is investigated and shows that the walk-off between the spatially multiplexed frequency-degenerated pump beams significantly limits the bandwidth of the conversion process for telecom applications. Finally, it is shown that the mode coupling and resulting interferential beating at the pump wavelength can impair the temporal profile of the converted signals. A strategy to overcome this deleterious effect is then proposed and based on a low-frequency non-degeneracy of the single pumping configuration.

## 2. PRINCIPLE-OF-OPERATION

The principle-of-operation of the FD-IFWM conversion process is displayed in Fig. 1(a). For the sake of simplicity, only two different spatial modes are here considered but the mechanism can be stacked and generalized for *n* higher-order modes. The same high-power continuous pump wave *P* (green arrow and circles) is simultaneously injected into two spatial modes 1 & 2 of a FMF, together with a weak signal *s* injected into the first mode (highlighted with red arrow and circle in Fig. 1(a)). The key point is that the frequency offset -$\Delta\omega$ of the signal *s* with respect to the pump frequency should be specifically adjusted in such a way that the relative group velocity (GV) evaluated at the average frequency of the two waves (*s* and *P*) in mode 1 is equal to that evaluated at the average frequency in the other mode (between *i* and *P*) [19-20]. Thanks to such a frequency-degenerated and spatially non-degenerated combination, a phase-matched IFWM interaction can be obtained resulting in the generation of an *idler* component *i* (blue arrow and circle) in the second spatial mode, symmetrically shifted from the pump frequency by +$\Delta\omega$. A simultaneous spatial and frequency conversion process of the signal wave is thus achieved by means of a single spatially-multiplexed pump wavelength. Note that this process can be also interpreted as a spatial generalization of the frequency-degenerated FWM interaction occurring in a birefringent fiber for which the two polarization modes correspond to mode 1 and 2 of the present case [30-31].

In terms of phase matching condition, the energy conservation law determines the corresponding *idler* frequencies $\omega_i$ in such a way that:

$$\omega_i = 2\omega_p - \omega_s, \quad (1)$$

where $\omega_p$ and $\omega_s$ are the pump and signal angular frequencies, respectively. While the momentum conservation for the FD-IFWM process between modes 1 and 2 leads to the phase mismatch $\Delta\beta$:

$$\Delta\beta = \beta^{(1)}(\omega_p) + \beta^{(2)}(\omega_p) - \beta^{(1)}(\omega_s) - \beta^{(2)}(\omega_i), \quad (2)$$

where $\beta^{(j)}$ denotes for the propagation constant in the spatial mode *j*. By developing the propagation constant of each mode in Taylor series around the pump frequency $\omega_p$ and neglecting higher-order dispersion effects above the second order, we get:

$$\Delta\beta = \Delta\omega \left[ \left(\beta_1^{(2)} + \frac{\Delta\omega}{2}\beta_2^{(2)}\right) - \left(\beta_1^{(1)} - \frac{\Delta\omega}{2}\beta_2^{(1)}\right) \right], \quad (3)$$

Where $\beta_1^{(j)}, \beta_2^{(j)}$ denote the inverse group velocity (IVG) and group velocity dispersion (GVD) at $\omega_p$ in mode *j*. This relation highlights the fact that the linear phase mismatch mainly depends on the difference of group velocity calculated at the central frequencies of each wave propagating in the different modes understudy. In order to satisfy this phase-matching condition, we then confirm from Eq. (3) that the group velocity evaluated at the average frequency propagating in both modes should be equal. It is also important to note, in particular for the section dealing with the influence of the walk-off effect, that this nonlinear process can be also interpreted in terms of group-velocity-matched inter-modal cross-phase modulation. Hence, the beating between pump and signal in mode 1 modulates the refractive index encountered by the continuous pump wave in mode 2, thus leading to the generation of a frequency sidebands at $\omega_i$ [33-34]. For this reason, in the following experimental section, the power involved in the fundamental mode will be always larger than the power injected into higher order modes, so as to maximize this nonlinear coupling.

Finally, the same principle can be generalized to *n* spatial modes by injecting the single frequency-degenerate pump wave in higher-order modes with phase-matched signals into the fundamental mode. To illustrate that *scenario*, Fig. 1(b) displays the case of the three higher-order modes of the 6-LP mode FMF involved in our experiment (see experimental section for a full description of the FMF). As can be seen on the experimental curves of the relative inverse group velocity (RIGV) [23], the degenerated pump wavelength (green vertical dashed-line) is fixed to 1565 nm for the $LP_{01}$, $LP_{11a}$, $LP_{02}$ and $LP_{31a}$ spatial modes, while the three input signals to be converted ($s_1$, $s_2$ and $s_3$) are injected into the fundamental mode with a wavelength offset close to 15 nm for a phase-matched *idler* generation in the $LP_{11a}$ mode, 24 nm for an *idler* generation in $LP_{02}$ and 27 nm for $LP_{31a}$. This is the reason why, in the experimental results described below, the signal wavelengths injected into $LP_{01}$ have to be tuned in the L-band (1565-1625 nm) so as to generate the corresponding *idler* waves in the C-band (1530-1565 nm) and thus match with the gain bandwidth of our available amplifiers. Note however that signal and idler wavelengths could be swapped in order to generate the idler wave in the L-band instead of the C-band. Finally, thanks to such dispersive curve measurements, the phase mismatch $\Delta\beta$ has been evaluated for each of the 10 modes of the FMF undertest in Fig. 1(c) and confirm the specific wavelength shift involved to satisfy the phase-matching condition. In addition, these curves highlight the fact that the degenerated spatial modes belonging to the same group of modes are characterized by a slightly different phase-matching condition. Therefore, in presence of random mode coupling, a broader but less efficient gain profile is expected to be observed.

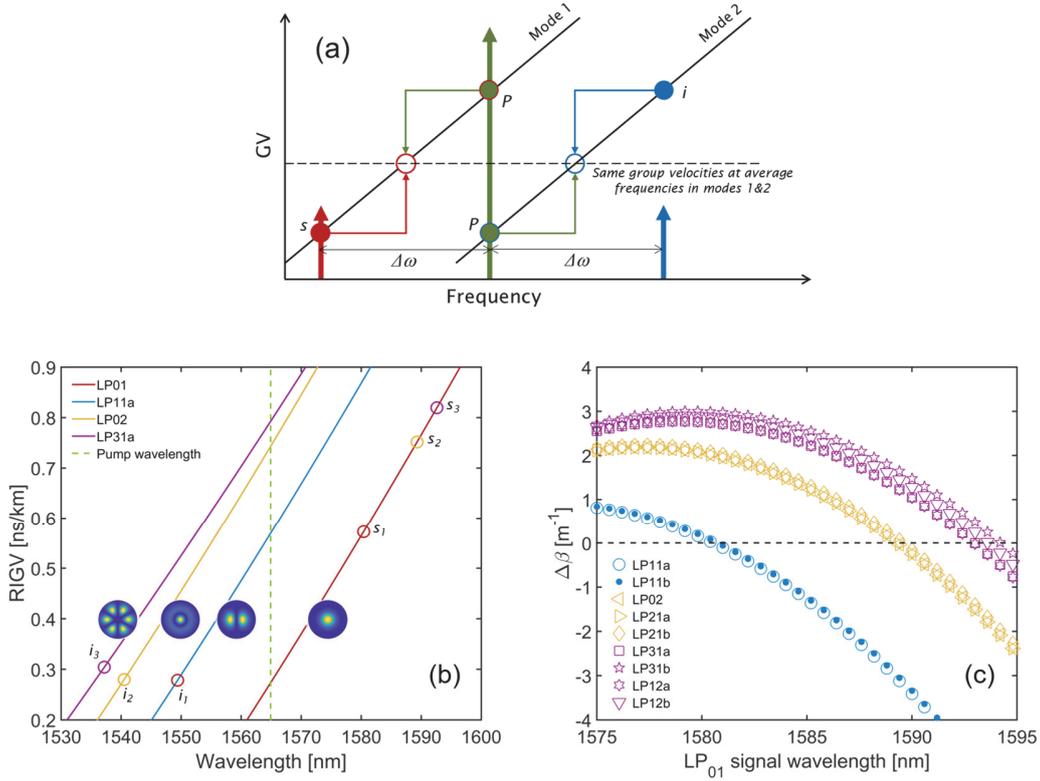

**Fig. 1.** (a) Principle-of-operation of the FD-IFWM process. (b) Experimental measurements of the relative inverse group velocity (RIGV) as a function of wavelength for each mode under study as well as illustration of the mode profile. (c) Corresponding phase-matching condition $\Delta\beta$ obtained from experimental data and Eq. (3) for all the nine higher-order spatial modes of our FMF.

## 3. EXPERIMENTAL SETUP

To demonstrate the simultaneous FD-IFWM modal and wavelength conversion process, we have implemented the experimental setup displayed in Fig. 2. Four tunable external cavity lasers are involved in this system. A first continuous pump wave (CW) centered at 1565 nm is phase-modulated by means of a $LiNBO_3$ phase modulator driven by a 100-MHz 29-dBm RF signal, in such a way to enlarge the pump spectral linewidth and limit the Brillouin back-scattering in the FMF undertest. The pump wave is then split into 4 different paths and amplified by means of a series of single-mode C-band 30-dBm Erbium-doped fiber amplifiers (EFDA) before simultaneous injection into the 1.8-km long 6-LP-mode fiber in respectively the $LP_{01}$, $LP_{11a}$, $LP_{02}$ and $LP_{31a}$ spatial modes. Note that these modes have been arbitrary chosen in each of the four spatial groups of modes of our FMF and similar results could be achieved in any other combination of degenerated spatial modes. Subsequently, three independent CW tunable signals in the L-band are combined together and amplified before being coupled to the fundamental mode. These signals can be intensity encoded thanks to a pulse pattern generator (PPG) driving a Mach-Zehnder intensity modulator for Telecommunication purpose and walk-off study. The different input pump waves as well as signals to convert are then coupled into the FMF by means of a 10-mode spatial multiplexer from Cailabs based on a multi-plane light conversion technology [35].

Polarization controllers are inserted into each optical input path to optimize the excitation efficiency of each spatial mode within the FMF. Indeed, due to random mode coupling between degenerated spatial modes as well as polarization dependent performance of the spatial multiplexer, we adjusted the input polarization to maximize the generation of the *idler* wave in each specific mode. The FMF consists in a 1.8-km long 6-LP-mode graded-index fiber (10 spatial modes including all degenerate modes) manufactured by Prysmian group with a core diameter of 22.5 μm [36]. The effective areas of each mode around 1550 nm are 75 μm$^2$ for the fundamental mode, 100 μm$^2$ for $LP_{11a}$ and $LP_{11b}$, 160 μm$^2$ for $LP_{02}$ and 170 μm$^2$ for $LP_{31a}$ and $LP_{31b}$. The dispersion curves of each mode shown in Fig. 1(b) have been measured by the time of flight method [23], which lead to a chromatic dispersion at the pump wavelength of 18.8 ps/nm/km for the fundamental mode, 19.3 ps/nm/km for $LP_{11a}$, 19.8 ps/nm/km for $LP_{02}$ and 18.3 ps/nm/km for $LP_{31a}$. The losses for all the modes are below 0.25 dB/km, and the maximum differential mode group delay between the modes is lower than 550 ps/km. More precisely, at the pump wavelength, the differential mode group delay between the fundamental mode and $LP_{11a}$ has been measured to 296 ps/km, 470 ps/km between $LP_{01}$ and $LP_{02}$ and 519 ps/km between $LP_{01}$ and $LP_{31a}$. At the output of the system, another polarization controller is implemented to apply mechanical stress on the FMF in order to maximize the energy at a specific output port of the spatial demultiplexer. Finally, a 10-port optical switch allows us to select one particular spatial mode at the output of the demultiplexer while a tunable filter is used to attenuate the leakages arising from any other spatial modes. The total losses of the system were measured at an average value of 9.3 dB for all the modes while the average cross-talks between the different groups of spatial modes due to the input multiplexer and linear mode coupling were found better than 20 dB (measured to 23 dB in average). Each spatial mode understudy can be then characterized in the spectral and temporal domains by means of an optical spectrum analyzer and an electrical sampling oscilloscope combined with a 12.5-GHz bandwidth photoreceiver.

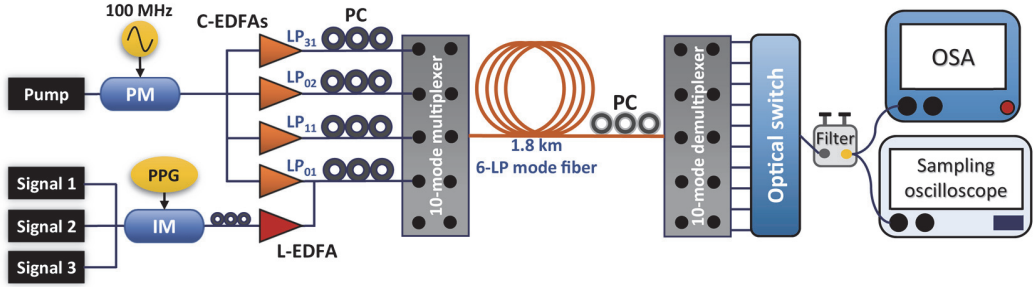

**Fig. 2.** Experimental setup. PM phase modulator, IM intensity modulator, PPG pulse pattern generator, PC polarization controller, EDFA Erbium-doped fiber amplifier, OSA optical spectrum analyzer.

## 4. EXPERIMENTAL RESULTS

### A. Phase-matching conditions

We have first experimentally and independently characterized the phase-matching conditions for all the three higher-order modes $LP_{11}$, $LP_{02}$ and $LP_{31}$. To this aim, we have mapped the FD-IFWM spectrum for fixed pump wavelength injected simultaneously into $LP_{01}$ and the higher-order mode undertest, while the signal wavelength injected into the fundamental mode is continuously swept. For each series of recordings, the pump wavelength is fixed to 1565 nm, while its injected power (measured at the input of the spatial multiplexer) is set to 33 dBm for $LP_{01}$ and 26 dBm for the corresponding higher-order mode. The signal power injected into the fundamental mode is adjusted 20 dB below the pump power. It is also important to notice that due to strong mode coupling in our FMF, the energy initially injected into one specific degenerated spatial mode is mainly spread in its whole corresponding group of modes. Therefore, to catch the whole dynamics of the FD-IFWM process, the resulting data are computed by combining all the spectra measured in each individual group of spatial modes. Hence, the resulting $LP_{11}$ spectrum corresponds to the sum of both demultiplexed $LP_{11a}$ and $LP_{11b}$ modes, $LP_{02}$ to the sum of the three modes $LP_{02}$, $LP_{21a}$ and $LP_{21b}$ while $LP_{31}$ corresponds to the sum of $LP_{31a}$, $LP_{31b}$, $LP_{12a}$ and $LP_{12b}$. Results are summarized in Figs. 3(a,d) for $LP_{11}$, Figs. 3(b,e) for $LP_{02}$ and Figs. 3(c,f) for $LP_{31}$. More precisely, Figs. (a-c) display from bottom to top, the concatenation of the output spectra recorded in a specific higher order mode as a function of the wavelength of the input signal injected into the fundamental mode, labeled $\lambda_{SLP01}$ along the vertical axes.

For all the three higher-order modes understudy, when the seed signal injected into the fundamental mode is swept (here the leakage is clearly visible in all the 3 modes), we can observe on the FD-IFWM spectral mapping the generation of an *idler* component in a specific range of $LP_{01}$ signal wavelengths $\lambda_{SLP01}$ (1580-1582 nm for $LP_{11}$, 1588-1593 nm for $LP_{02}$ and 1590-1597 nm for $LP_{31}$), corresponding to the phase-matching condition, in good qualitative agreement with the phase mismatch curves of Fig. 1(c). Note that sweeping the signal wavelength out of this specific band rapidly decreases the efficiency of the process. The bottom line displays the corresponding individual spectrum obtained at the maximum of efficiency for each group of modes (optimum phase-matching condition). These results confirm that both spatial and wavelength conversion phenomenon can be achieved in each group of modes thanks to a single frequency-degenerated pumped IFWM process.

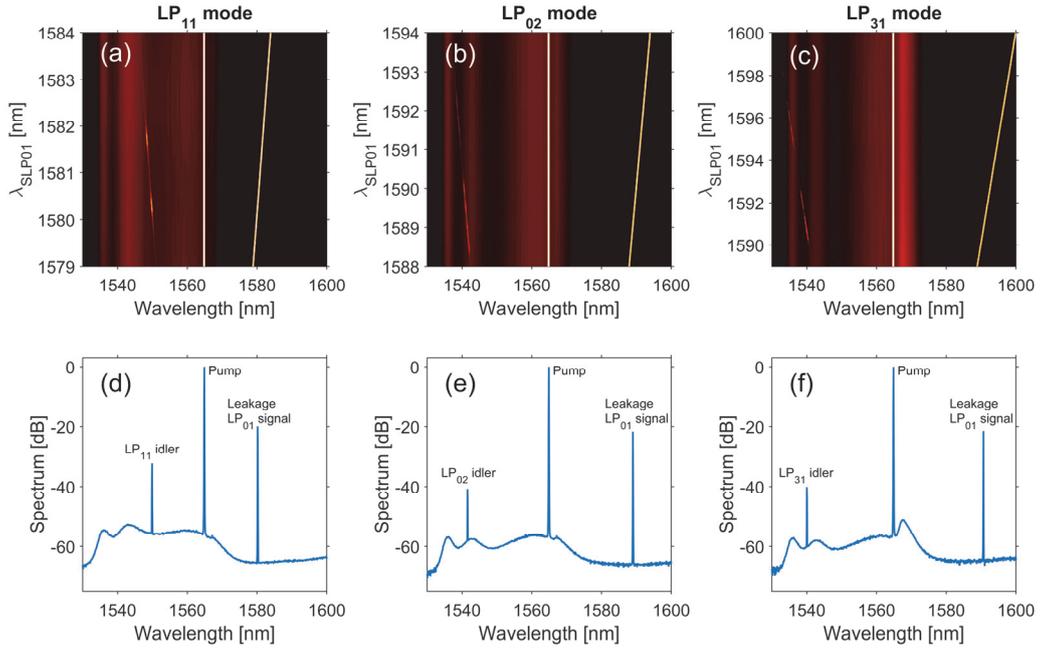

**Fig. 3.** Experimental spectral mapping of the FD-IFWM process. (a-c) Top-line corresponds to the output spectra in each group of modes as a function of the signal wavelength $\lambda_{SLP01}$ injected into the fundamental mode when pumping the two modes with a single 1565-nm pump wavelength. (d-f) Optimum spectrum obtained for $LP_{11}$ (d), $LP_{02}$ (e) and $LP_{31}$ (f), respectively.

### B. Conversion efficiency

In order to further illustrate and quantify the different bandwidths of the FD-IFWM process, we have reported for each group of modes shown in Figs. 3, the CE as a function of the signal wavelength injected in $LP_{01}$ ($\lambda_{SLP01}$). These curves confirm the high sensitivity of the process with respect to the linear phase-matching condition illustrated here by a CE bandwidth close to 1 nm for each mode understudy. We also note the multiple peaks nature of the CE curves, for instance two well-distinct maxima for $LP_{11}$ in Fig. 4(a). We attribute this behavior to the degeneracy of the spatial modes undertest which are characterized by a different phase-matching condition due to a slightly different group-velocity, as already shown in Fig. 1(c). Note that the sensitivity of this process could be then exploited to characterize such a weak difference. More importantly, the decrease of the CE with respect to the number of degenerate modes belonging to the same modal group appears particularly relevant in Fig. 4. For instance, the CE of the modal group corresponding to $LP_{31}$ (4 spatial-modes, 8 in polarization) is decreased by more than 10 dB compared to $LP_{11}$. Indeed, for larger group of modes, the CE is then blurred by the random mode coupling between adjacent and degenerate spatial modes as well as polarization fluctuations [24-27]. Moreover, the nonlinear efficiency is also reduced due to a larger mode area and larger mode-coupling area.

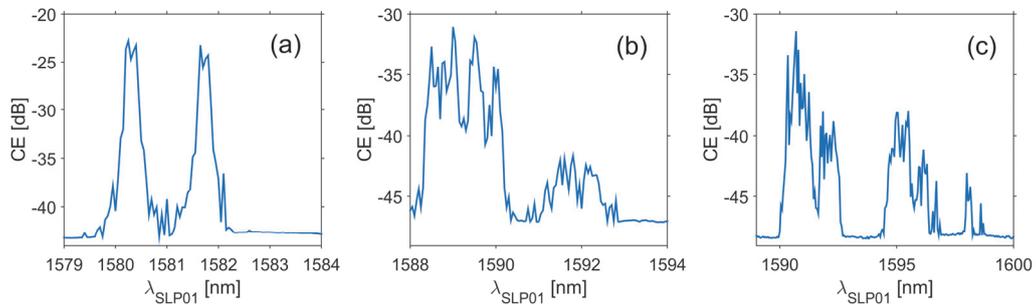

**Fig. 4.** Conversion efficiency of the FD-IFWM process for each group of modes as a function of signal wavelength $\lambda_{SLP01}$ injected in $LP_{01}$. (a) $LP_{11}$, (b) $LP_{02}$ and (c) $LP_{31}$.

## C. Temporal characterization

In order to further characterize the FD-IFWM process and its potential application for dual spatial/wavelength conversion operations, we have replaced the CW seed-signal injected into the fundamental mode by a Non-Return-to-Zero (NRZ) 1-Gbit/s Telecom signal. We then monitored the resulting output eye-diagram of the converted signal for the $LP_{11}$ spatial mode. The wavelength of the seed-signal in $LP_{01}$ is adjusted to be as close as possible to the optimum phase-matching condition, around 1580.25 nm, while the pump wavelength injected in $LP_{01}$ and $LP_{11a}$ is fixed to 1565 nm. Figure 5(a) depicts the temporal profile of the output $LP_{11}$ *idler* wave obtained for pump powers of 33-dBm and 17-dBm in the fundamental mode and $LP_{11}$, respectively. We can clearly observe a huge amount of amplitude jitter which completely closes the resulting eye-diagram. This phenomenon originates from the interferential beating occurring between the frequency-degenerated pump wave injected into the $LP_{11}$ and the leakage resulting from $LP_{01}$ due to spatial multiplexing imperfection (∼20 dB of extinction). Indeed, even if the two pump beams originate from the same initial laser, the different paths of the experimental setup as well as external perturbations are sufficient to scramble the phase difference between the two pump waves. Consequently, for moderate pump power injected into the $LP_{11}$ mode, both waves have approximatively the same energy and interfere with a maximum of contrast, leading to a total closure of the eye-diagram.

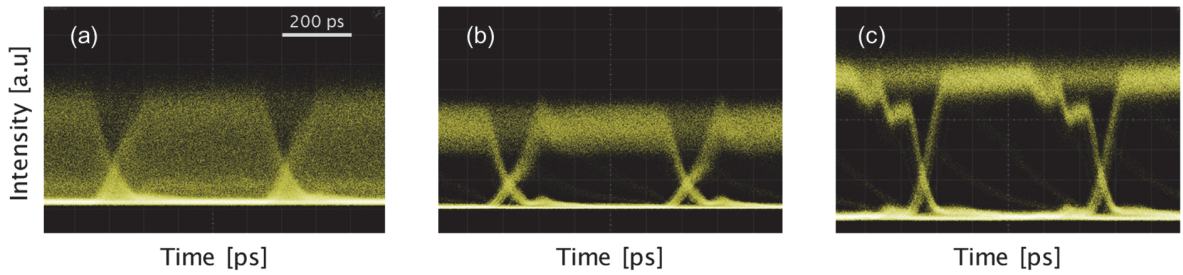

**Fig. 5.** Temporal characterization of the converted $LP_{11}$ *idler* wave as a function of pump configuration when a 1-Gbit/s telecom signal is injected into the fundamental mode. (a) 17-dBm pump power injected into the $LP_{11}$ mode, 33-dBm in $LP_{01}$. (b) 26-dBm pump power injected in $LP_{11}$, 33-dBm in $LP_{01}$. (c) 26-dBm low-frequency degenerated 30-GHz offset pump wave injected into the $LP_{11}$, 33-dBm in $LP_{01}$. The 200-ps solid-line segment indicates the time-scale for the 3 eye-diagrams.

In order to overcome this deleterious effect, two *scenarii* have been tested. The first one consists in increasing the pump power injected into the $LP_{11}$ mode so that the $LP_{01}$ leakage could be considered as negligible, thus reducing the amplitude of the interference pattern. This solution is illustrated in Fig. 5(b) for which the 26-dBm pump signal injected into the $LP_{11}$ mode is 10 times stronger than the $LP_{01}$ pump leakage. We can still observe the presence of a non-negligible amount of amplitude jitter on the top of the NRZ signal, but with a strong reduction compared to the previous case. Finally, the last option, displayed in Fig. 5(c) consists in a low-frequency non-degeneracy of the pump wave, for which a slightly frequency-offset second laser, is exploited to pump the higher-order mode. The frequency shift of this dual-pump configuration has to be adjusted to be above the receiver bandwidth, in such a way for the photoreceiver to be blind for this beating, as well as sufficiently close to the $LP_{01}$ pump wave frequency to still satisfy the phase-matching condition of the FD-IFWM. Figure 5(c) shows the resulting eye-diagram of the output $LP_{11}$ *idler* wave when a 30-GHz offset is applied for the $LP_{11}$ 26-dBm pump wave. We can observe that the interferential amplitude beating is almost suppressed from the temporal profile, leading to a wide opened eye-diagram for the converted signal. However, the drawback is that one has to pump the FMF with two different CW lasers, at the cost of simplicity, instead of one single pump for the classical FD-IFWM. Nevertheless, a third option would be to inject the pump wave by means of a free-space setup in order to avoid the temporal decorrelation and subsequently manage the ratio of energy incoming into each group of spatial modes.

## D. Walk-off effect

Thanks to the 30-GHz low-frequency non-degenerated dual-pump configuration described above, we have then characterized the FD-IFWM conversion process in terms of bandwidth for Telecom applications. To this aim, we have implemented the modal/wavelength conversion process of a NRZ signal from the fundamental mode to the $LP_{11a}$ spatial mode. The pump wavelength is still fixed to 1565 nm for both spatial modes $LP_{01}$ and $LP_{11a}$, while the $LP_{01}$ NRZ seed-signal is set to 1580.25 nm in accordance with previous measurements and phase-matching condition. Pump and signal powers are the same as in the previous experiment. The bit-rate of the seed-signal injected into the $LP_{01}$ is then gradually increased from 1 Gbit/s to 10 Gbit/s. Figure 6 summarizes the recordings of the output eye-diagrams after spatial demultiplexing and spectral filtering of the converted *idler* wave. While the 1-Gbit/s converted signal, shown in Fig. 6(a), displays a wide opened eye-diagram, we can observe a strong distortion of the temporal profile, in particular a large pulse broadening, with respect to the repetition-rate increase, especially for 10-Gbit/s in Fig. 6(c). This behavior can be well understood by recalling that the FD-IFWM process can be interpreted as a cross-phase modulation

(XPM) between a high-power modulated pump wave propagating along the fundamental mode and a continuous wave injected into a higher-order mode. Note that it mainly explains why the pump power injected into the fundamental mode is larger than in higher order modes, so as to maximize this XPM process. Moreover, since these two spatially multiplexed signals propagate at different speeds in the FMF due to DMGD, the XPM interaction is not group-velocity matched and a walk-off effect is expected to blur the temporal profile for high repetition-rate signals. Regarding the $LP_{11a}$ present case, the DMGD accumulated along the fiber length between the $LP_{01}$ modulated pump wave and the corresponding $LP_{11}$ CW pump signal is estimated to 270 ps according to the experimental measurements of Fig. 1(b). Consequently, this amount of DMGD clearly limits the maximum bit-rate of the input signal to less than 2 or 3 Gbit/s, in good agreement with the output eye-diagrams depicted in Figs. 6.

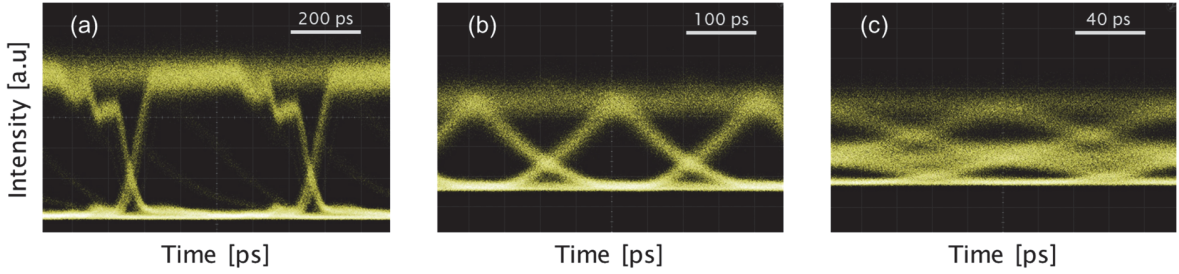

**Fig. 6.** Eye-diagram of the output $LP_{11a}$ *idler* wave as a function of the bit-rate of the seed signal injected into the fundamental mode. Experimental parameters are the same as in Fig. 5(c). (a) 1 Gbit/s. (b) 5 Gbit/s. (c) 10 Gbit/s.

**E. Threefold spatial and wavelength conversion of 1-Gbit/s signals**

Finally, to underline the potential of multiple parametric conversions in FMF for Telecom applications, we have performed a parallel threefold conversion operation at 1 Gbit/s from the fundamental mode towards the three higher-order modes understudy, i.e. $LP_{11a}$, $LP_{02}$ and $LP_{31a}$. To this aim, all the higher-order spatial modes undertest are simultaneously pumped around 1565 nm with a 30-GHz frequency offset pump wave compared to the $LP_{01}$ mode (low-frequency non-degenerate double pump scheme). The pump powers involved in this proof-of-principle experiment are set to 33 dBm for the fundamental mode and 26 dBm for all the three higher-order modes. Three independent signals are then NRZ encoded at 1-Gbit/s and coupled to the fundamental modes. The different wavelengths are adjusted in such a way to fulfill the phase-matching condition for each targeted high-order mode, i.e. 1580.25 nm for $LP_{11a}$, 1589.7 nm for $LP_{02}$ and 1590.82 nm for $LP_{31a}$. At the output of the system, each converted *idler* wave is spatially demultiplexed, filtered out thanks to a programmable liquid-crystal-on-silicon based optical filter, amplified by means of a C-band EDFA and subsequently characterized in the time domain. Figure 7(a) reports the full optical spectrum recorded at the output of the FMF. Note that this resulting total spectrum has been computed by combining all the spectra measured in each individual port of the output spatial demultiplexer. This spectrum shows that converted *idler* waves ($i_1^{11}$, $i_2^{02}$ and $i_3^{31}$) are generated in each higher-order mode in accordance with the three initial signals injected into the fundamental mode ($s_1^{01}$, $s_2^{01}$ and $s_3^{01}$). Figures 7(b)-7(d) display the output eye-diagrams for all the demultiplexed and filtered 1-Gbit/s *idler* waves. Note that to optimize one particular eye-diagram in any degenerate spatial mode, the input polarization of its corresponding pump beam and output stress on the FMF before the spatial demultiplexer have to be optimized in such a way to avoid interference induced amplitude jitter between spatially degenerate modes of the same group. For all the resulting converted signals, the corresponding eye-diagrams appear quite opened. However, a non-negligible amount of amplitude jitter can be observed for the highest order modes, which is attributed to the low CE in these particular modes due to low nonlinearity and random mode coupling. However, this proof-of-principle experiment confirms the possibility to achieve a parallel multi-signal spatial and wavelength conversion process based on FD-IFWM.

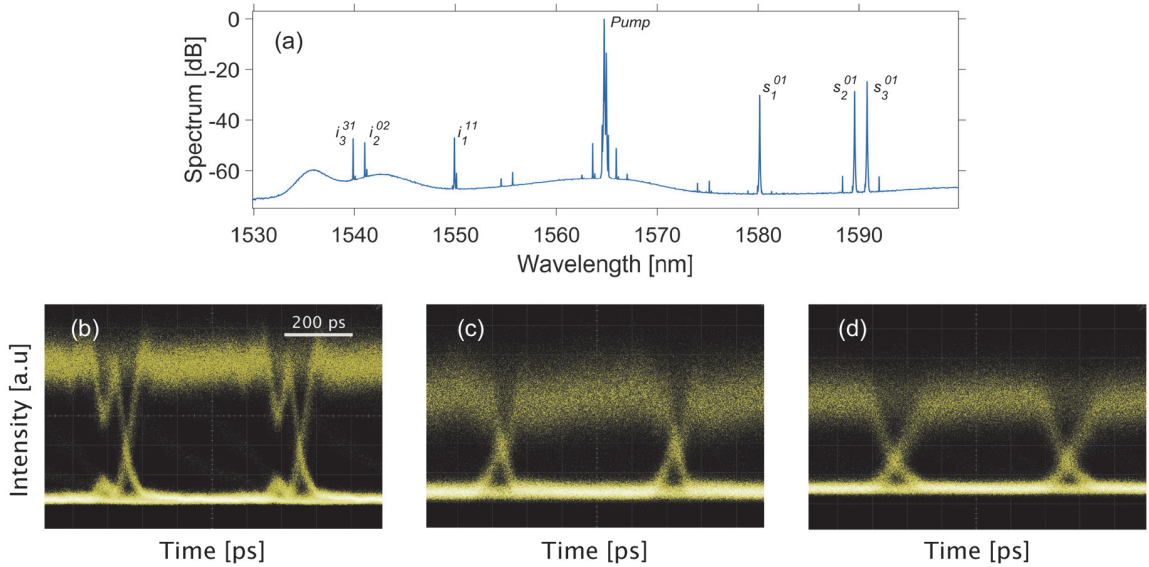

**Fig. 7.** (a) Total spectrum recorded at the output of the FMF for the threefold spatial and wavelength conversion process of 1-Gbit/s signals. $i_n^m$ corresponds to the $n^{th}$ *idler* wave generated in mode *m* due to the FD-IFWM process between the pump wave *Pump* and signal $s_n^{01}$ injected into the fundamental mode. (b) Eye-diagram of the output converted $LP_{11a}$ *idler* wave. (c) Eye-diagram of the $LP_{02}$ *idler* wave (d) Eye-diagram of the $LP_{31a}$ *idler* wave.

## 5. CONCLUSION

To conclude, we have experimentally demonstrated a simultaneous threefold wavelength and modal conversion process of 1-Gbit/s Telecom signals in the C+L band thanks to a 1.8-km long 6-LP-mode graded-index few-mode fiber. The principle-of-operation is based on phase-matched frequency-degenerated inter-modal four-wave mixing (FD-IFWM) phenomena occurring between the fundamental mode and the targeted higher-order modes. To this aim, a high-power pump wavelength is simultaneously injected into the $LP_{01}$, $LP_{11a}$, $LP_{02}$ and $LP_{31a}$ modes of our graded-index FMF together with three independent signals in the fundamental mode. Thanks to phase-matched FD-IFWM mechanisms, the initial signals are then simultaneously frequency and spatially converted from the fundamental mode to specific high-order modes. The temporal characterizations show that the converted signals can provide good quality eye-diagrams at 1 Gbit/s. However, our study also reveals the physical limits of such nonlinear conversion process, which could significantly impact its potential application for Telecom purpose. First-of-all, the differential modal group delay between the different signals induces a non-negligible walk-off between the spatially multiplexed signals and fundamentally limits the bandwidth of the conversion process to a few Gbit/s. Secondly, the performance of such a system are limited by the weak conversion efficiency of the FD-IFWM process due to the absence of parametric gain at the signal frequency, compared to the normal dispersion regime [14], and due to the weak nonlinearity involved by the large effective areas of the spatial modes. Linear mode coupling between higher-order spatially degenerated modes, as well as random polarization fluctuations, are also found to limit the long-term stability of our system [24-27]. Furthermore, the limited extinction ratio between the spatially multiplexed pump waves as well as the lack of coherence between these signals appears as a non-negligible source of impairments at the receiver. We have found that this drawback can be overcome thanks to the implementation of a low-frequency degenerated dual-pump configuration, but at the cost of simplicity. Further improvements could be achieved, in particular the bandwidth of the FD-IFWM process, thanks to a specific design of the FMF, with lower chromatic dispersion, normal dispersion regime at pump wavelength and smaller effective modal areas as well as the breaking of degeneracy for higher-order modes by means of elliptical core few-mode fibers [28, 37-38]. Note also that similar results could be demonstrated in a step-index FMF and could be generalized to higher numbers of spatial modes. However, since the DMGD is much larger in step-index FMF, longer wavelength conversion shifts could be achieved but at the cost of the bandwidth due to larger walk-off effects. In conclusion, these proof-of-principle experiments demonstrate that nonlinear functionalities, such as multiple spatial-wavelength conversion process, can be performed in FMFs and open the way to new all-optical processing techniques for spatial division multiplexing applications.

**Funding.** This work was funded by the Agence Nationale pour la recherche ANR APOFIS project ANR-17-ERC2-0020-01.

**Acknowledgment.** This work benefits from the PICASSO platform in ICB. The authors thank Dr. P. Béjot and T. Sylvestre for fruitful discussions.